\documentclass[reprint, showpacs, superscriptaddress]{revtex4-1}

\usepackage{graphicx, hyperref}

\usepackage{CJK}

\begin{document}

\begin{CJK*}{UTF8}{}

\title{Excited-state entanglement and thermal mutual information in random spin chains}

\CJKfamily{gbsn}

\author{Yichen Huang (黄溢辰)}
\email{yichenhuang@berkeley.edu}
\affiliation{Department of Physics, University of California, Berkeley, Berkeley, California 94720, USA}

\author{Joel E. Moore}
\affiliation{Department of Physics, University of California, Berkeley, Berkeley, California 94720, USA}
\affiliation{Materials Sciences Division, Lawrence Berkeley National Laboratory, Berkeley, California 94720, USA}

\date{\today}

\begin{abstract}
Entanglement properties of excited eigenstates (or of thermal mixed states) are difficult to study with conventional analytical methods.  We approach this problem for random spin chains using a recently developed real-space renormalization group technique for excited states (``RSRG-X'').  For the random $XX$ and quantum Ising chains, which have logarithmic divergences in the entanglement entropy of their (infinite-randomness) critical ground states, we show that the entanglement entropy of excited eigenstates retains a logarithmic divergence while the mutual information of thermal mixed states does not.  However, in the $XX$ case the coefficient of the logarithmic divergence extends from the universal ground-state value to a universal interval due to the degeneracy of excited eigenstates.  These models are noninteracting in the sense of having free-fermion representations, allowing strong numerical checks of our analytical predictions. 
\end{abstract}


\maketitle

\end{CJK*}

Concepts from quantum information theory have been widely used in condensed matter and atomic physics \cite{AFOV08, ECP10} to characterize quantum correlations in various interesting classes of states.  One such concept is quantum entanglement \cite{PV07, HHHH09}, which for critical ground states~\cite{holzhey, VLRK03, LRV04, LR09, CC04, CC09}, topological phases~\cite{Kitaev06b, Levin05, LH08}, and Fermi liquids~\cite{wolf, klich} provides unique insights into the physics that are difficult to obtain via other quantities.  Entanglement is also quantitatively related to the difficulty of describing one-dimensional (1D) noncritical \cite{Has07} and critical (``finite-entanglement scaling'' \cite{TOIL08, pollmann, PhysRevB.86.075117}) ground states by matrix product wave functions \cite{FNW92, PVWC07} in numerical approximations \cite{VC06, SWVC08}.

In this Rapid Communication, we study random spin chains, where entanglement is known to capture important aspects of the ground state~\cite{RM04, RM09}, and examine how the entanglement of individual excited eigenstates is different from the mutual information of thermal mixed states at nonzero temperature. This question can be viewed as an entanglement version of the classical problem of equivalence of ensembles: Is the canonical ensemble described by the density matrix $\rho=\exp(-H/T)$, where $H$ is the Hamiltonian and $T$ is the temperature, equivalent for important observables to the microcanonical ensemble of energy eigenstates with the same energy density? As $\rho$ is a mixed state, we need a notion that generalizes the entanglement entropy (well defined only for pure states), and mutual information (though not an entanglement measure) is a commonly used option.

Another motivation for studying excited eigenstates in random spin chains is the high level of current interest~\cite{gornyimirlin,basko:2006,Oganesyan:2007,palhuse,reichman, bardarson,kjall, voskaltman,RE14, PRA+14,serbyn, chandran} in how disorder (modeled by randomness) can lead to localized states violating the eigenstate thermalization hypothesis, even in the presence of interactions; this phenomenon is known as many-body localization. The eigenstate thermalization hypothesis~\cite{Deutsch:1991,Srednicki:1994,Rigol:2008} is that (for some not yet delineated classes of quantum many-body systems) local measurements of an energy eigenstate approach those of the thermal mixed state with the same energy density.  Intuitively, one region of the system sees the rest of the system as a bath or reservoir capable of providing energy and particles.  Localization does not support the transport of energy or particles and hence prevents full thermalization. We emphasize that many-body localization is a property associated with all eigenstates (not just the ground state) of disordered systems.

Excited eigenstates are ``physical'' states participating in the dynamics of the system, and hence their singularities strongly suggest a dynamical quantum phase transition. For example, in the random quantum Ising chain we find that the entanglement of (almost) all eigenstates becomes singular (i.e., diverges logarithmically) at the critical point. This is indeed accompanied with a dynamical quantum phase transition characterized by the time evolution of entanglement entropy \cite{RE14}.

The real-space renormalization group (RSRG)~\cite{MaDas1979,MaDas1980,Fis92,Fis94, Fis95,monthusreview} is a standard technique for ``infinite-randomness'' ground states in random spin chains. It has recently been generalized to excited states with the acronym RSRG-X~\cite{PRA+14}. Adapting this approach to our context, we make analytical predictions for the scaling of excited-state entanglement (defined as the average entanglement entropy of energy eigenstates sampled from a canonical ensemble) and thermal mutual information (the mutual information of a thermal mixed state) in the random $XX$ and quantum Ising chains, which are verified numerically.

We find that excited-state entanglement and thermal mutual information behave very differently. The latter behaves as one might expect for physical quantities at nonzero temperatures above a (random) quantum critical point: The characteristic divergence~\cite{RM04} is cut off by temperature. The former retains such a divergence, i.e., the entanglement entropy of excited eigenstates diverges logarithmically as that of the ground state does. There is a surprise: In the random $XX$ chain, the coefficient of the logarithmic divergence extends from the universal ground-state value to a universal interval due to the degeneracy of excited eigenstates (it is basis dependent and is determined only after a way of lifting the degeneracy of excited eigenstates is given).

\emph{Preliminaries.} We start by introducing key definitions and then review RSRG.  Entanglement reflects a remarkable fact about the product structure of the Hilbert space for a bipartite quantum system $AB$.  This Hilbert space is constructed as the tensor product of the Hilbert spaces for the two subsystems, i.e., it is spanned by product states made from (basis) vectors of $A$ and $B$.  However, the superposition principle allows linear combinations of product states, and in general such a linear combination is not a product of any wave functions in $A$ and $B$.

The entanglement entropy of a pure state $\rho_{AB}$ is the von Neumann entropy $S(\rho_A)=-\mathrm{tr}\rho_A\ln\rho_A$ of the reduced density matrix $\rho_A=\text{tr}_B\rho_{AB}$. It is the standard measure of entanglement for pure states. For mixed states, there are some entanglement measures in the literature and no single one is standard~\cite{PV07, HHHH09}. Most of these entanglement measures reduce for pure states to entanglement entropy, and are difficult (NP-hard \cite{Hua14}) to compute. Quantum mutual information $I(\rho_{AB})=S(\rho_A)+S(\rho_B)-S(\rho_{AB})$ is not an entanglement measure, as it is generically nonzero for separable (i.e., unentangled) states. It quantifies the total (classical and quantum) correlation between $A$ and $B$ in a possibly mixed state $\rho_{AB}$, and is the quantum analog of mutual information (the standard measure of correlation between two random variables) in classical information theory.

Let $S_L(|\psi\rangle)$ be the entanglement entropy of the state $|\psi\rangle$ in a spin model, where $A$ consists of a block of $L$ spins, and by default $|\psi\rangle$ is the ground state. $S_L$ satisfies an area law \cite{ECP10} in 1D gapped systems \cite{Has07}. In 1D gapless systems, $S_L\sim (c\ln L)/3$ \cite{holzhey, CC04, CC09} if the critical theory is a conformal field theory with central charge $c$, e.g., $S_L\sim (\ln L)/3$ in the homogeneous $XX$ and antiferromagnetic Heisenberg chains \cite{VLRK03, LRV04, LR09}. Similarly, let $I_L^T$ be the mutual information of the thermal mixed state $\exp(-H/T)$ at nonzero temperature $T$. $I_L^T$ always satisfies an area law \cite{WVHC08, scalingdiscord}, regardless of the energy gap or the dimension (geometry) of the lattice.

\emph{Real-space renormalization group.} As a standard analytical approach to the low-energy physics in random spin chains, RSRG is successful in practice and believed to be asymptotically exact at infinite-randomness quantum critical points. We briefly illustrate this approach in the context of the random $XX$ chain \cite{Fis94}. See Ref. \cite{monthusreview} and references therein for details and more examples.

The Hamiltonian is $H=\sum_iH_i$ with $H_i=J_i(\sigma_x^i\sigma_x^{i+1}+\sigma_y^i\sigma_y^{i+1})$, where $J_i$'s are independent and identically distributed (i.i.d.) random variables. At each step of RSRG, we find the strongest bond $J_j=\max_iJ_i=:\Omega$ and diagonalize $H_j$. Assuming $J_j\gg J_{j\pm1}$, the spins $j$ and $j+1$ form a singlet (the ground state of $H_j$), and then degenerate perturbation theory (Schrieffer-Wolff transformation \cite{BDL11}) leads to an effective interaction
\begin{equation} \label{eff}
J_{j-1,j+2}=J_{j-1}J_{j+1}/J_j<\Omega
\end{equation}
between the spins $j-1$ and $j+2$ (the last row in Table \ref{diag}). As such, we eliminate the strongest bond $J_j$ and reduce the energy scale $\Omega$. Repeating these steps, the ground state of the random $XX$ chain is approximately a tensor product of singlets. Moreover, Eq. (\ref{eff}) induces a RSRG flow equation for the distribution of $J_i$'s. There is a simple infinite-randomness fixed point solution as the attractor for all nonsingular initial distributions of $J_i$'s \cite{Fis94}, which justifies the assumption $J_j\gg J_{j\pm1}$ in the asymptotic limit. Therefore, the low-energy physics of the random $XX$ chain is universal: It is governed by the fixed point distribution, regardless of initial distributions.

\begin{table}
\caption{Eigenvalues and eigenstates of $H_j$; effective interactions $H_{j-1,j+2}=J_{j-1,j+2}(\sigma_x^{j-1}\sigma_x^{j+2}+\sigma_y^{j-1}\sigma_y^{j+2})$.}
\begin{tabular}{ccc}
\hline \hline
Eigenvalues&Eigenstates&Effective interactions\\
\hline
$2J_j$&$|\uparrow_j\downarrow_{j+1}\rangle+|\downarrow_j\uparrow_{j+1}\rangle$&$J_{j-1,j+2}=J_{j-1}J_{j+1}/J_j$\\
0&$|\uparrow_j\uparrow_{j+1}\rangle,~|\downarrow_j\downarrow_{j+1}\rangle$&$J_{j-1,j+2}=-J_{j-1}J_{j+1}/J_j$\\
$-2J_j$&$|\uparrow_j\downarrow_{j+1}\rangle-|\downarrow_j\uparrow_{j+1}\rangle$&$J_{j-1,j+2}=J_{j-1}J_{j+1}/J_j$\\
\hline \hline
\end{tabular}
\label{diag}
\end{table}

The entanglement entropy $S_L$ is proportional to the number of singlets across one boundary of the block \cite{RM04}. Let $\Gamma=\ln(\Omega_0/\Omega)$ with $\Omega_0$ the initial energy scale. The RSRG flow equation and the fixed point distribution imply (a) $\lambda\sim\Gamma^2$, where $\lambda$ is the length scale of the singlets at the energy scale $\Omega$, and (b) $N\sim(\ln\Gamma)/3$, where $N$ is the average total number of singlets across a particular cut at energy scales greater than $\Omega$. Substituting $\lambda\sim L$,
\begin{equation} \label{RM}
\langle S_L\rangle\sim2N\ln2\sim(\ln2)(\ln L)/3,
\end{equation}
where $\langle\cdot\rangle$ denotes averaging over randomness. See Refs. \cite{RM04, RM09} for details.

\emph{Numerics.} In free-fermion systems, the algorithm for computing entanglement entropy is well established \cite{PE09}. It is used in Refs. \cite{VLRK03, LRV04, Laf05} to compute the entanglement entropy of ground states in the homogeneous (and random) $XX$ chain, quantum Ising chain, etc., and it also works for excited eigenstates. The algorithm for computing the mutual information of thermal states is a variant of it \cite{PE09}. Technically, these algorithms make use of (i) the fact that a free-fermion system can be decomposed into a bunch of noninteracting fermionic modes, and (ii) the observation that the eigenstates and the thermal states of a free-fermion Hamiltonian are (fermionic) Gaussian states, i.e., they can be reconstructed from their covariance matrices. Since these algorithms are efficient in the sense that their running time grows polynomially with the system size, we are able to simulate chains of 200--1000 spins with a laptop and extract the coefficient of ``$\ln L$'' convincingly. We verify with accurate numerics all implications of RSRG and RSRG-X for the scaling of excited-state entanglement and thermal mutual information. This is a numerical test of the recently developed RSRG-X \cite{PRA+14}.

\emph{Excited-state entanglement in random $XX$ chain.} Let $\{|\psi_i\rangle\}$ be a complete set of eigenstates of $H$, and define
\begin{equation} \label{def}
S_L^T=\frac{\sum_i\exp(-\langle\psi_i|H|\psi_i\rangle/T)S_L(|\psi_i\rangle)}{\sum_i\exp(-\langle\psi_i|H|\psi_i\rangle/T)}
\end{equation}
as the average entanglement entropy of eigenstates $|\psi_i\rangle$'s sampled from the Boltzmann distribution at temperature $T$. Here $T$ is a parameter tuning the (average) energy, for we wonder whether (and how) the scaling of excited-state entanglement depends on energy. Alternatively, one may study the average entanglement entropy of eigenstates with (close to) a particular energy. Note that $S_L^T$ is not the entanglement of the thermal mixed state $\exp(-H/T)$.

RSRG-X \cite{PRA+14} is an approach to the long-range physics of excited states in random spin chains. Following our previous discussion of RSRG, we show implications of RSRG-X for the scaling of $\langle S_L^T\rangle$ in the random $XX$ chain.

At each step of RSRG-X, we still diagonalize $H_j$: The eigenvalues and eigenstates are given in Table \ref{diag}. Here the spins $j$ and $j+1$ are in a random eigenstate of $H_j$ sampled from the Boltzmann distribution at temperature $T$ (cf. they are always in the ground state of $H_j$ in RSRG), and then degenerate perturbation theory leads to an effective interaction between the spins $j-1$ and $j+2$: Different eigenstates may induce different interactions, but fortunately the difference is only in sign (the right column in Table \ref{diag}). Hence the flow equation and the fixed point solution for the distribution of $|J_i|$'s in RSRG-X are identical to those in RSRG.

We calculate the amount of entanglement generated in RSRG-X. (i) If the spins $j$ and $j+1$ are in an eigenstate of $H_j$ with eigenvalue $\pm2J_j$ (the first or the last row in Table \ref{diag}), then a unit of entanglement is generated as in RSRG. (ii) Otherwise, the spins $j$ and $j+1$ may be in a superposition of $|\uparrow\uparrow\rangle$ and $|\downarrow\downarrow\rangle$ (the middle row in Table \ref{diag}), and an undetermined amount of entanglement is generated. Let $\alpha$ be the ratio of the amount of entanglement generated in RSRG-X to that generated in RSRG. Averaging cases (i) and (ii) gives $1/2\le\alpha\le1$. At any constant temperature $T>0$, the energy scale $\Omega$ becomes much lower than $T$ after some number of RSRG-X steps. Hence the scaling of $\langle S_L^{T>0}\rangle$ is the same as that of $\langle S_L^{T=\infty}\rangle$. Summarizing,
\begin{equation} \label{main}
\langle S_L^{T=\infty}\rangle\sim\langle S_L^{T>0}\rangle\sim\alpha(\ln2)(\ln L)/3,~1/2\le\alpha\le1.
\end{equation}
It is not a limitation of our approach that the prefactor $\alpha$ is undetermined. Indeed, the eigenvalues of a random $XX$ Hamiltonian are degenerate. Hence the complete set of eigenstates $\{|\psi_i\rangle\}$ and the scaling of $\langle S_L^T\rangle$ (\ref{def}) are not unique. We construct two examples in which $\alpha=1/2$ and $\alpha=1$, respectively.

\begin{figure}
\includegraphics[width=\linewidth]{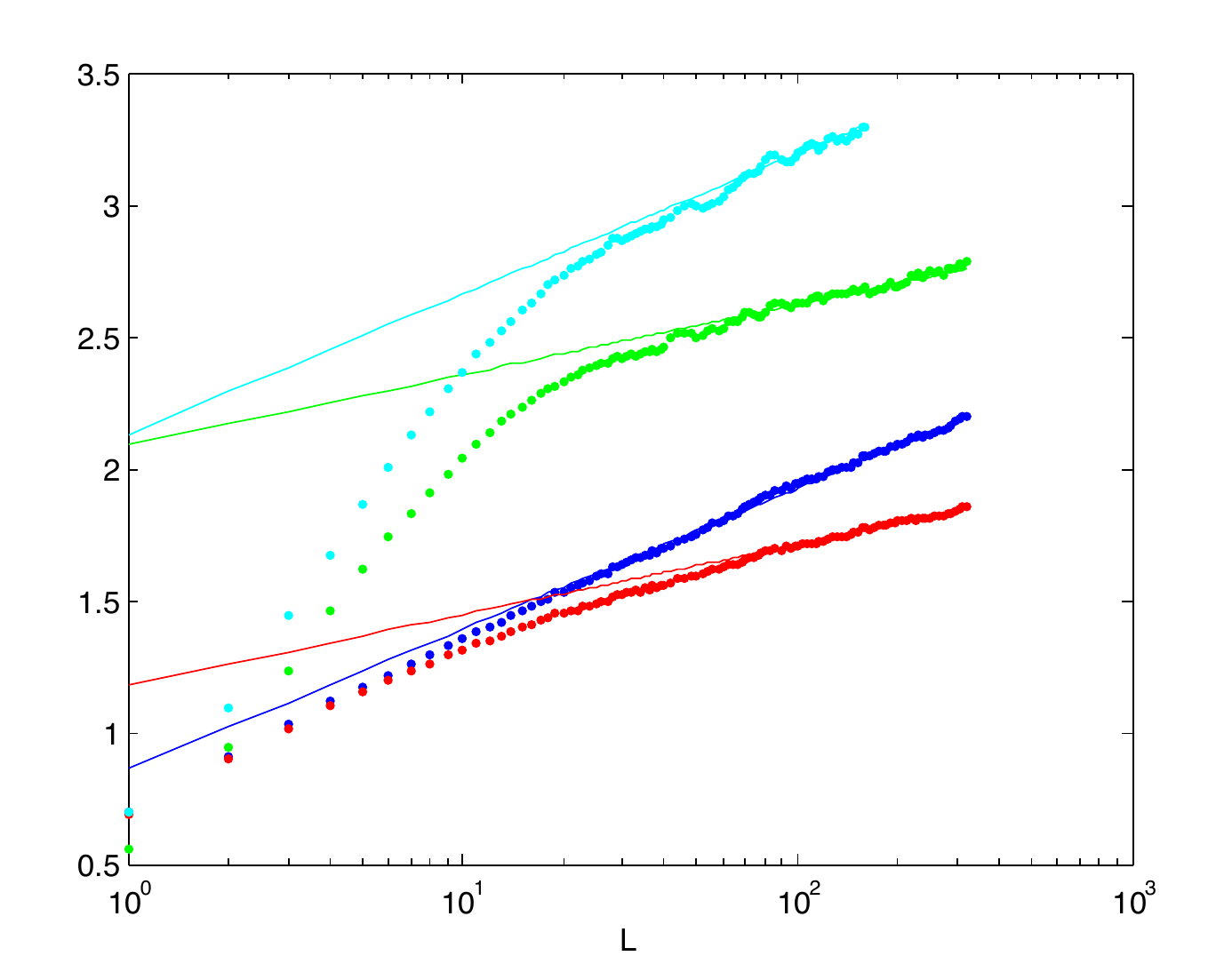}
\caption{(Color online) Scaling of $\langle S_L^T\rangle$ in random $XX$ chain, which is apparently different from the entanglement scaling of some excited states in homogeneous spin chains \cite{alba}. Here $J_i$'s are i.i.d. uniform random variables on the interval $[0,1]$. Example 1 ($\alpha=1/2$). The blue, red, and green dots are data (averaged over $5000$ samples) for $T=0,10^{-3}$, and $\infty$, respectively, in chains of $1000$ spins. The lines are fits based on (\ref{RM}) and (\ref{main}): $(\ln2)(\ln L)/3+0.86$ (blue), $(\ln2)(\ln L)/6+1.18$ (red), and $(\ln2)(\ln L)/6+2.09$ (green). Example 2 ($\alpha=1$). The cyan dots are data (averaged over 3000 samples) for $T=\infty,~\delta=10^{-9}$ in chains of $500$ spins. The cyan line is a fit based on (\ref{main}): $(\ln2)(\ln L)/3+2.13$.}
\label{1}
\end{figure}

\emph{Example 1.} Since the total magnetization $\sigma_z=\sum_i\sigma_z^i$ is conserved, one may require that each $|\psi_i\rangle$ is an eigenstate of $\sigma_z$, which is physically interpreted as fixing the fermion number in the fermion representation. Then, in case (ii) the spins $j$ and $j+1$ are (approximately) in either $|\uparrow\uparrow\rangle$ or $|\downarrow\downarrow\rangle$ (not  a superposition) so that (almost) no entanglement is generated. Hence $\alpha=1/2$, which is verified numerically for $T=\infty$ (green) and $T=10^{-3}$ (red) in Fig. \ref{1}. Note that the (universal) logarithmic scaling starts at larger $L$ for $T=10^{-3},\infty$ than for $T=0$ (blue).

\emph{Example 2.} Let $H'=\sum_i(1+\delta)J_i\sigma_x^i\sigma_x^{i+1}+J_i\sigma_y^i\sigma_y^{i+1}$ such that $\lim_{\delta\rightarrow0}H'=H$. The eigenvalues of $H'$ are generically nondegenerate. Then, in case (ii) the spins $j$ and $j+1$ are (approximately) in $(|\uparrow\uparrow\rangle\pm|\downarrow\downarrow\rangle)/\sqrt2$ (maximally entangled state) so that (almost) one unit of entanglement is generated. Hence $\alpha=1$, which is verified numerically for $T=\infty$ (cyan) in Fig. \ref{1}.

\begin{figure}
\includegraphics[width=\linewidth]{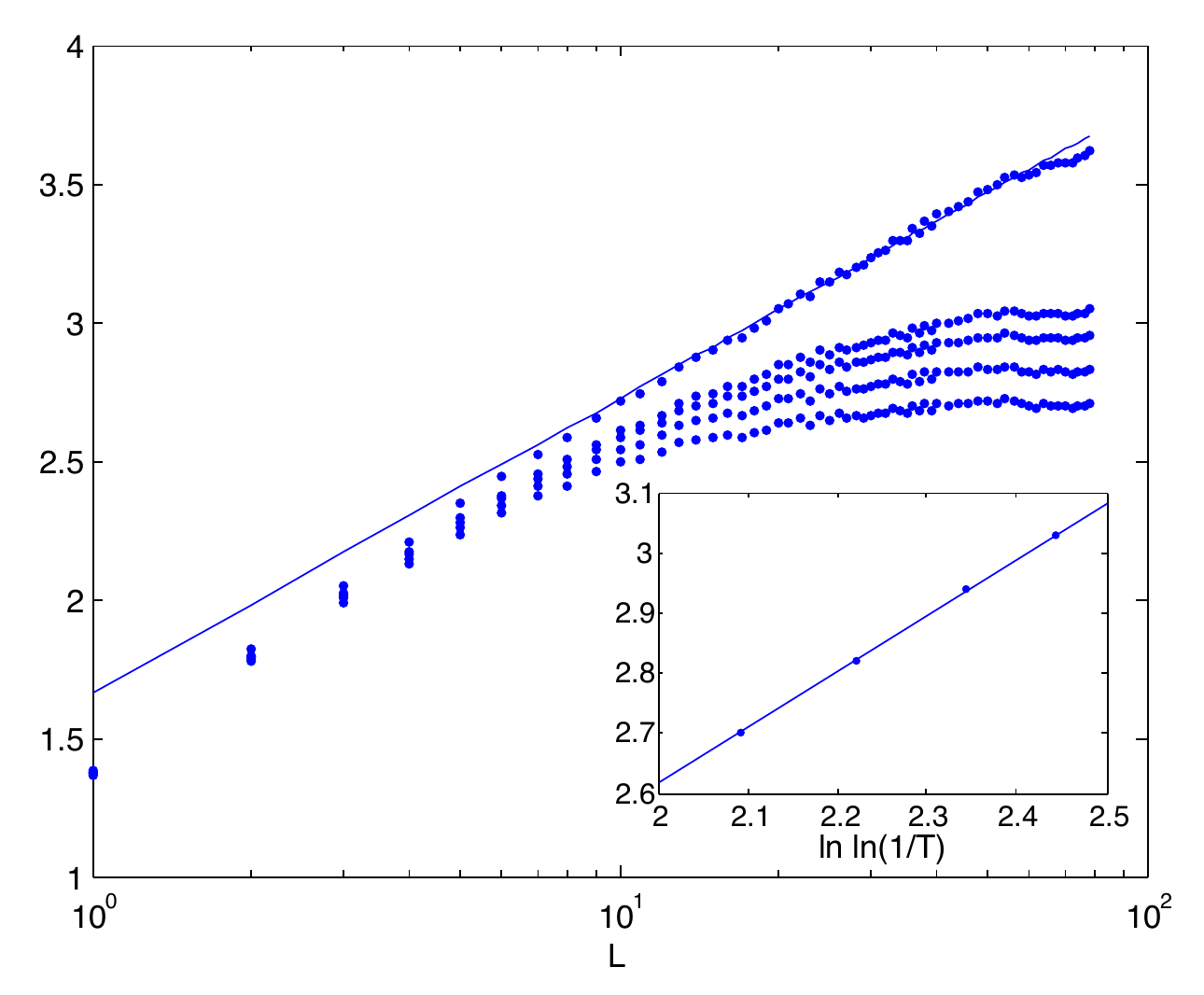}
\caption{(Color online) Scaling of $\langle I_L^T\rangle$ in random $XX$ chain. The dots (from top to bottom) are data (averaged over $2500$ samples) for $T=0,10^{-5},3\times10^{-5},10^{-4}$, and $3\times10^{-4}$ in chains of $200$ spins. The line is a fit based on (\ref{T}): $2(\ln2)(\ln L)/3+1.66$. $\langle I_L^T\rangle$ behaves as if $T=0$ for $L\ll L_c$ and saturates for $L\gg L_c$. Inset: Saturation value $\langle I_{L\gg L_c}^T\rangle$ vs temperature. The line is a fit based on (\ref{T}): $4(\ln2)[\ln\ln(1/T)]/3+0.77$.}
\label{2}
\end{figure}

\emph{Thermal mutual information in random XX chain.} We calculate the scaling of $\langle I_L^T\rangle$ using RSRG. (i) If $\Omega\gg T$, we do RSRG as if $T=0$. (ii) If $\Omega\ll T$, the remaining spins are in the maximally mixed state as if $T=\infty$. (iii) The transition occurs at $\Omega_c\sim T$:
\begin{equation}
L_c\sim\Gamma_c^2\sim\ln^2(1/\Omega_c)\sim\ln^2(1/T).
\end{equation}
The thermal mixed state $\exp(-H/T)$ is approximately of the form $\rho_1\otimes\rho_0$, where $\rho_1$ is a tensor product of singlets, and $\rho_0$ is a maximally mixed state. Hence,
\begin{eqnarray} \label{T}
I_L^T&\approx& I_L(\rho_1\otimes\rho_0)=I_L(\rho_1)+I_L(\rho_0)=I_L(\rho_1)\Rightarrow\nonumber\\
\langle I_L^T\rangle&\sim&\langle I_{L_c}^T\rangle\sim2(\ln2)(\ln L_c)/3\sim4(\ln2)[\ln\ln(1/T)]/3,\nonumber\\
\langle I_L^T\rangle&\sim&\langle I_L^{T=0}\rangle\sim2(\ln2)(\ln L)/3
\end{eqnarray}
for $L\gg L_c$ and $L\ll L_c$, respectively, or compactly
\begin{equation}
\langle I_L^T\rangle\sim2(\ln2)[\ln\min\{L,\ln^2(1/T)\}]/3,
\end{equation}
which is verified numerically in Fig. \ref{2}.

Any entanglement measure (for mixed states) satisfies the following: (a) It does not increase under local operations and classical communication (LOCC); (b) it reduces to entanglement entropy for maximally entangled states; and (c) other postulates irrelevant to us. See Refs. \cite{PV07, HHHH09} for details on the axiomatic approach to entanglement measures. Since the states $\rho_1$ and $\rho_1\otimes\rho_0$ can be transformed to each other by LOCC,
\begin{eqnarray}
E_L^T:&=&E_L(\exp(-H/T))\approx E_L(\rho_1\otimes\rho_0)=E_L(\rho_1)\nonumber\\
&=&S_L(\rho_1)=I_L(\rho_1)/2\approx I_L^T/2\Rightarrow\langle E_L^T\rangle\sim\langle I_L^T\rangle/2
\end{eqnarray}
for any entanglement measure $E$ (including, but not limited to, entanglement cost, distillable entanglement, entanglement of formation, relative entropy of entanglement, squashed entanglement, and logarithmic negativity). Note that logarithmic negativity, while it does not reduce to entanglement entropy for all pure states, does reduce to entanglement entropy for maximally entangled states and hence satisfies the postulate (b) above. We do not expect any of the aforementioned entanglement measures can be computed efficiently even in free-fermion systems.

\begin{figure}
\includegraphics[width=\linewidth]{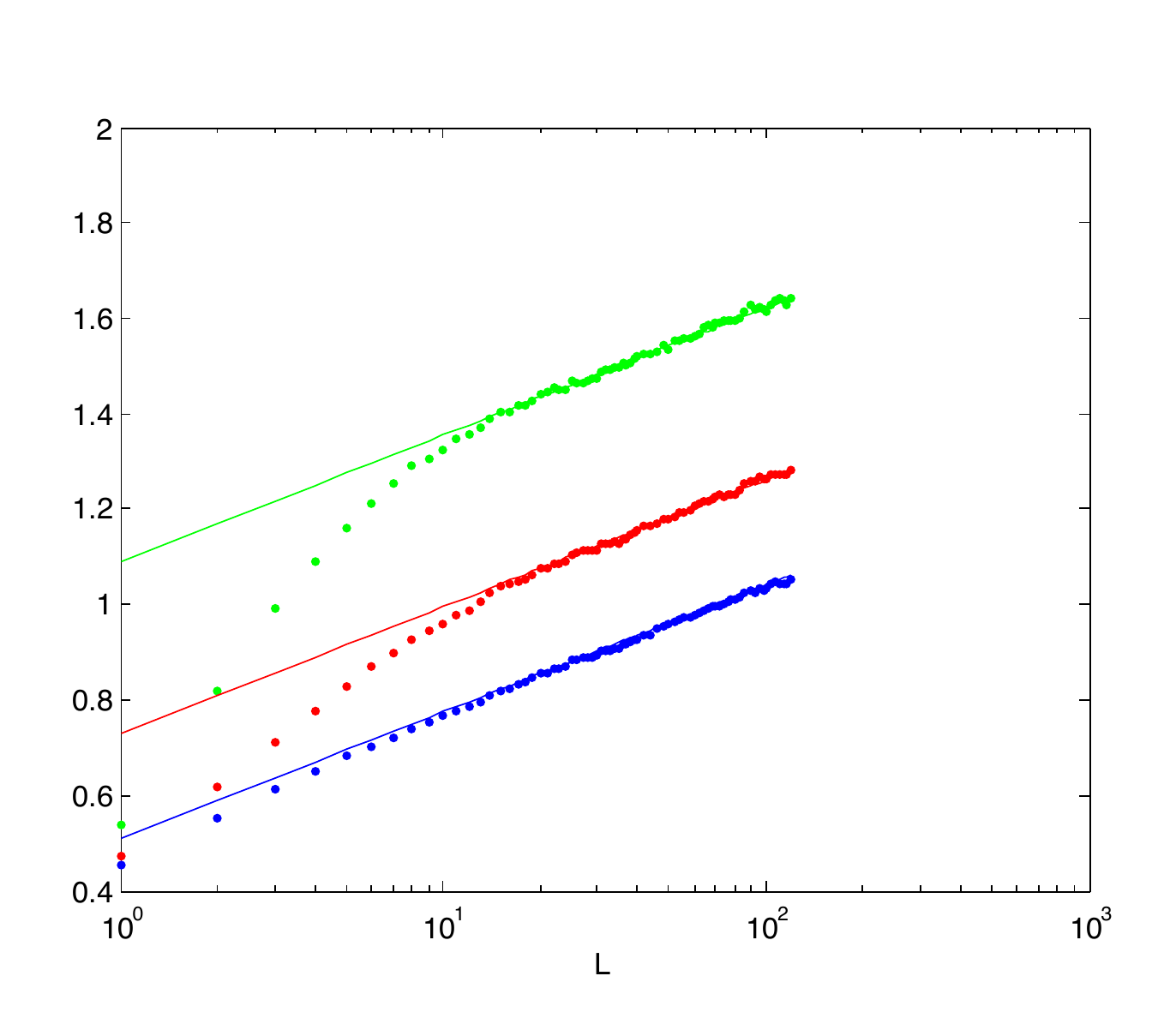}
\caption{(Color online) Scaling of $\langle S_L^T\rangle$ in critical random quantum Ising chain ($\delta=0$). Here $J_i,h_i$'s are i.i.d. uniform random variables on the interval $[0,1]$. The blue, red, and green dots are data (averaged over $2500$ samples) for $T=0,0.2$, and $\infty$, respectively, in chains of $400$ spins. The lines are fits based on (\ref{I}): $(\ln2)(\ln L)/6+0.51$ (blue), $(\ln2)(\ln L)/6+0.73$ (red), and $(\ln2)(\ln L)/6+1.09$ (green).}
\label{3}
\end{figure}

\emph{Random quantum Ising chain.} We now study the random quantum Ising chain \cite{Fis92, Fis95, monthusreview}. The Hamiltonian is
\begin{equation} \label{Ising}
H=\sum_iJ_i\sigma_x^i\sigma_x^{i+1}+h_i\sigma_z^i,
\end{equation}
where $J_i$'s are i.i.d. and $h_i$'s are i.i.d. random variables. The eigenvalues of $H$ are generically nondegenerate. Let $\delta=(\overline{\ln|h|}-\overline{\ln|J|})/(\mathrm{var}\ln|h|+\mathrm{var}\ln|J|)$. At $\delta=0$, the system is critical, and RSRG implies \cite{RM04}
\begin{equation} \label{I}
\langle S_L^{T=0}\rangle\sim(\ln2)(\ln L)/6.
\end{equation}
Otherwise ($\delta\neq0$) we expect an area law for $\langle S_L^{T=0}\rangle$. Let $\xi\sim1/\delta^2$ be the characteristic length scale within and beyond which the system appears critical and noncritical, respectively \cite{Fis95}. The saturation value is $\langle S_{L\gg\xi}^{T=0}\rangle\sim(\ln2)(\ln\xi)/6\sim(\ln2)(\ln|1/\delta|)/3$ for $|\delta|\ll1$. Straightforward perturbative calculations show that fortunately the difference between effective interactions induced in RSRG and RSRG-X is only in sign \cite{PRA+14}. Hence the flow equation and the fixed point solution for the distributions of $|J_i|,|h_i|$'s in RSRG-X are identical to those in RSRG. Moreover, the amount of entanglement generated in RSRG-X is the same as that generated in RSRG. Therefore $\langle S_L^{\forall T}\rangle\sim\langle S_L^{T=0}\rangle$, which is verified numerically in Fig. \ref{3}.

\emph{Beyond free-fermion systems.} Consider the weakly interacting model $H'=H+\sum_iJ'_i\sigma_z^i\sigma_z^{i+1}$, where $H$ is the random quantum Ising Hamiltonian (\ref{Ising}), and $J'_i$'s ($\ll J_i,h_i$'s) are i.i.d. random variables. This model is studied using RSRG-X in Ref. \cite{PRA+14}: There is strong numerical evidence for a temperature-tuned dynamical quantum phase transition. After developing intuitions about this transition, the scaling of entanglement will be clear.

Irrelevant perturbations do not change the universality class of phase transitions, but they modify the strength of relevant terms. In RSRG-X, the $J'$ perturbations are irrelevant \cite{PRA+14}. Let $\delta_r(\delta,T,J')$ be the ``renormalized $\delta$,'' which is a function of $T$ because the implementation of RSRG-X is temperature dependent. The critical temperature $T_c$ is given by $\delta_r(\delta,T_c,J')=0$. Therefore,
\begin{equation}
\langle S_L^{T=T_c}\rangle\sim(\ln2)(\ln L)/6.
\end{equation}
We expect an area law for $\langle S_L^{T\neq T_c}\rangle$, and the saturation value is $\sim(\ln2)[\ln|1/(T-T_c)|]/3$ for $|T-T_c|\ll1$ (and finite $T_c$).

\emph{Note added.} Recently, we became aware of a paper \cite{RRS14} that studies the entanglement of states with a small finite number of excitations. It should be clear that we have studied the entanglement of states with a finite energy density above the ground state, i.e., an infinite number of excitations in the thermodynamic limit.

\emph{Acknowledgment.} This work was supported by DARPA OLE (Y.H.) and NSF DMR-1206515 (J.E.M.).

\end{document}